\documentclass[aps, prd, reprint, longbibliography, nofootinbib,superscriptaddress, floatfix]{revtex4-1}

\usepackage{verbatim}
\usepackage[T1]{fontenc}
\usepackage{mathbbol}
\usepackage[dvipsnames]{xcolor}
\usepackage{orcidlink}

\usepackage[normalem]{ulem}
\usepackage[english]{babel}
\usepackage{physics}
\usepackage{graphicx,color,mathtools}
\usepackage{amsthm,amsmath,amssymb,mathrsfs}
\usepackage{bm,bbm}
\usepackage{float}
\definecolor{myred}{RGB}{179, 27, 27}
\usepackage{hyperref}
\hypersetup{
    colorlinks=true,
    linkcolor=myred, 
    citecolor=myred, 
    urlcolor=myred  
 }

\newcommand{\CI}{\text{Ci}}
\newcommand{\calR}{\mathcal{R}}

\begin{document}

\title{Nonlinear Dynamics near the Threshold of Gravitational Collapse}

\author{Jaime Redondo-Yuste\orcidlink{0000-0003-3697-0319}}
\email[]{jredondo@jhu.edu}
\affiliation{William H. Miller III Department of Physics and Astronomy,
Johns Hopkins University, 3400 North Charles Street, Baltimore, Maryland, 21218, USA}

\author{Josu C. Aurrekoetxea\orcidlink{0000-0001-9584-5791}}
\affiliation{Center for Theoretical Physics, Massachusetts Institute of Technology, Cambridge, MA 02139, USA}

\begin{abstract}
Perturbation theory is an essential tool to model and interpret gravitational dynamics, for example, binary black hole mergers. Therefore it is also crucial to precisely understand its regimes of validity. The collapse of a scalar field under its own self-gravity provides a clean laboratory to study these questions. By varying the field's initial amplitude we can transition smoothly between a perturbative regime, where the field scatters in an approximately flat spacetime; and a nonperturbative regime, where a black hole forms in finite time. In this work, we use numerical relativity simulations of this set-up to investigate the accuracy of a perturbative expansion around flat spacetime, including next-to-next-to-leading order effects. Our simulations show deviations from these perturbative predictions before black hole formation, once the maximum luminosity of the process is sufficiently large, $\mathcal{L}_{\rm peak} \sim 10^{-2} \mathcal{L}_{\rm Planck}$. We characterize these nonlinear effects including a redshift of the driving frequency and a power-law spectral amplitude, which we show is consistent with approximate discrete self-similarity. These results provide a step forward towards understanding the limits of perturbative expansions in more realistic strong-gravity phenomena such as non-spherical collapse and high-velocity black hole mergers. 
\end{abstract}

\maketitle

\section{Introduction}

Perturbation theory is one of the most powerful tools of modern science~\cite{Batterman:2001}. In essence, it allows us to trade the study of complex, nonlinear, and asymmetric systems for linear deformations of exact, highly symmetric solutions. Remarkably, it is often effective beyond the regime where we may expect it to hold. Examples of this are known in quantum electrodynamics~\cite{Dyson:1952tj}, quantum chromodynamics~\cite{tHooft:1973alw}, and also to describe the two-body problem in General Relativity~\cite{Buonanno:1998gg, Buonanno:2006ui,vandeMeent:2020xgc}.  

In the gravitational context, the effectiveness of perturbative methods is remarkable. A variety of approximations are routinely applied to describe the dynamics of binary black hole (BH) mergers, including post-Newtonian ($v/c \ll 1$), post-Minkowskian ($GM \ll 1$), multipolar, small mass ratio or self-force ($m_1m_2 \ll m_1+m_2$), and ringdown ($|g_{\mu\nu}-g^{\rm Kerr}_{\mu\nu}|\ll 1$) approximations~\cite{Blanchet:1985sp,Buonanno:1998gg,Will:2011nz,Blanchet:2013haa,Damour:2016gwp,Damour:2017zjx,Bern:2019nnu,Berti:2025hly, Pound:2021qin, Wardell:2021fyy}. Although calibration to numerical relativity simulations is indispensable to provide waveforms that are accurate enough for data-analysis purposes, the accuracy of methods that rely \emph{only} on (resummations of) perturbation theory is astounding~\cite{Buonanno:1998gg, Damour:2001tu, vandeMeent:2020xgc}. Key to this is the flexibility of considering perturbations around a flat spacetime (e.g. post-Minkowskian expansion), around a single BH spacetime (self-force or ringdown), or around an effective geometry (e.g., effective-one-body~\cite{Buonanno:1998gg, Damour:2001tu}). Whether a problem is (non)perturbative obviously depends on the background spacetime under consideration. 

There are (at least) three instances where perturbation theory fails to capture the nonlinear dynamics of BHs\footnote{We restrict our discussion to four dimensions and asymptotically flat BHs. A richer phenomenology takes place in higher dimensions or asymptotically anti-de-Sitter spacetimes.}: (i) the onset of BH formation~\cite{Choptuik:1992jv}, (ii) the nonlinear saturation of instabilities of rapidly rotating BHs~\cite{East:2013mfa, Yang:2014tla}, and (iii) grazing collisions and scattering of BHs, compact objects, or pulses of light at ultrarelativistic speeds~\cite{East:2012mb, Cook:2017fec,Zhu:2026mhn, Patino:2026jlq}. In some of these regimes~\cite{Yang:2014tla}, where trapping is sufficiently efficient, or the system is externally driven~\cite{Ma:2025rnv}, gravity can become turbulent, signaling a strong breakdown of perturbation theory around single BH backgrounds. This implies that the success of the aforementioned methods at characterizing the GW emission of BH mergers may break down in some corners of parameter space. We posit that a key open question is precisely to quantitatively characterize the regime where perturbative methods around known, exact solutions of Einstein's equations fail, and General Relativity enters a strongly nonlinear and dynamical regime. 

We begin tackling this question by considering the simplest possible system that admits a clean perturbative description, as an expansion in a parameter $\epsilon$ and that shows a departure from this perturbative description as $\epsilon \to 1$. This is achieved by the gravitational collapse of a massless, spherically symmetric scalar field, first studied mathematically by Christodoulou~\cite{Christodoulou:1986zr}, and later numerically by Choptuik~\cite{Choptuik:1992jv}. In this case, by letting $\epsilon$ control the amplitude of the initial scalar pulse, we transition from a \emph{dispersive} regime, where the scalar field propagates along an approximately flat spacetime; to a \emph{collapsing} regime, where the scalar field promptly forms a BH. The problem of critical collapse has received significant attention in the past three decades, extending to different matter fields and beyond spherical symmetry~\cite{Garfinkle:1994jb, Evans:1994pj, Abrahams:1993wa, Abrahams:1994nw,Gundlach:1995kd,Choptuik:1996yg, Gundlach:1997wm,Garfinkle:1998va,Choptuik:2003ac, Choptuik:2004ha, Hilditch:2013cba,Baumgarte:2019fai,Baumgarte:2023tdh, Marouda:2024epb,Baumgarte:2026pqr}. For a comprehensive review, we refer the interested reader to~\cite{Gundlach:2025yje}.

Our main result is diagnosing the breakdown of perturbation theory around flat spacetime before the onset of gravitational collapse, and characterizing the nonlinear dynamics in this regime. This breakdown occurs (perhaps surprisingly) \emph{very} close to the threshold of collapse $\epsilon \lesssim 1$, and is identified by the breakdown of the (perturbative) scaling of several diagnostic quantities with $\epsilon$. We also observe strongly nonlinear effects that we cannot explain by a perturbative analysis around flat space, though we note that these may possibly be explained by resummations of perturbation theory, or alternatively by perturbations around the critical solution, i.e., an expansion in $|\epsilon-1| \ll 1$ rather than in $\epsilon \ll 1$. This includes, in particular, a flattening of the spectrum of the outgoing scalar field, which transitions from being an approximately discrete spectrum, to continuous. We later show that this flattening, including in particular the exponent of its power-law, can be predicted if we assume that the solution is approximately discretely self-similar close to the threshold of collapse. 

The flattening of the spectrum, as well as other deviations from perturbation theory around flat spacetime, is only visible close to the threshold of collapse. This suggests a sort of ``nonlinear cosmic censorship'', by which horizon formation hides strongly nonlinear features~\cite{Bhagwat:2017tkm} from external observers. We show that not all the nonlinear effects are visible to observers at infinity. For instance, higher harmonics are excited, but they decay as $r^{-2}$. Therefore they do not contribute to any net flux at future null infinity -- only observers that get sufficiently close to the ``strong-field'' region will be able to measure them~\cite{Cardoso:2026llh}. This is unlike nonlinear harmonics excited during the black hole ringdown, which are observable~\cite{Cheung:2022rbm, Mitman:2022qdl, Lagos:2022otp}. Closely related is the ``maximum luminosity'' conjecture~\cite{Deruelle:1984hq,Dyson:2023, Gibbons:2002iv, Barrow:2014cga, Barrow:2017atq, Cardoso:2013krh,Cardoso:2018nkg, Jowsey:2021gny}, which postulates that no physical process can achieve a luminosity larger than Planck's luminosity $\mathcal{L}_P = c^5/G \approx 3.6 \times 10^{52}\mathrm{W}$. Our results suggest that the dimensionless maximum luminosity is a good proxy to quantify how ``nonlinear'' a gravitational process is: here we achieve maximum luminosities of about $\mathcal{L} \sim 7 \times 10^{-2}\mathcal{L}_P$. Critical collapse simulations in~\cite{Cardoso:2018nkg} and more recently in~\cite{Patino:2026jlq} lead to luminosities as large as $\mathcal{L} \sim 0.2 \mathcal{L}_P$ -- for comparison, a binary black hole merger like GW150914 only achieves a maximum luminosity $\mathcal{L} \approx 10^{-3} \mathcal{L}_P$~\cite{LIGOScientific:2016vlm}. 

Perturbative methods become accurate once again when $\epsilon > 1$, and at late enough times following black hole formation. In that regime, the behavior of the scalar field is governed by the \emph{ringdown} of a scalar field scattering off a Schwarzschild BH. In this way, this problem admits multiple perturbative regimes around different background spacetimes, in clear analogy with BH coalescences.

The paper is structured as follows: Section~\ref{sec:numerics} briefly describes the set-up and the numerical methods employed to evolve the system, followed by section~\ref{sec:perturbation}, which summarizes certain key results from a perturbative analysis. Our results are presented in Section~\ref{sec:results}, and we discuss our findings and future research directions in section~\ref{sec:conclusions}. In the following, we work with geometric units $G=1=c$ and with the mostly plus $(-+++)$ signature.

\section{Set-Up {\em \&} Numerics}\label{sec:numerics}

We consider a massless real scalar field $\Phi$, evolving in a spherically symmetric geometry $g$. The dynamics is governed by the Einstein-Klein-Gordon equations
\begin{equation}\label{eq:ekg}
    \begin{aligned}
        G_{ab}[\Phi] =& 8\pi \Bigl[\nabla_a\Phi\nabla_b\Phi - \frac{1}{2}g_{ab}\nabla_c\Phi\nabla^c\Phi\Bigr] \, , \\
        \square_g \Phi =& 0 \, .
    \end{aligned}
\end{equation}
We prescribe static initial data at some spacelike slice $\Sigma_0$ given by 
\begin{align}\label{eq:ics}
    \Phi\vert_{\Sigma_0} &= \epsilon A_* \cos(\omega_0 r) e^{-(r-r_0)^2/(2\sigma^2)} \, , \\
    \partial_t\Phi\vert_{\Sigma_0} &= 0 \, , 
\end{align}
where $\epsilon,\omega_0, r_0, \sigma$ parametrize the initial data family, $r$ is the spherical (radial) coordinate, and $A_*$ is the critical amplitude, i.e., the amplitude such that initial data with $\epsilon < 1$ disperses, while initial data with $\epsilon > 1$ collapses and forms BHs~\cite{Christodoulou:1986zr,Choptuik:1992jv}. Therefore, $\epsilon < 1 (>1)$ corresponds to sub-critical or dispersive (super-critical or collapsing) solutions. While the numerical solution with exactly $\epsilon = 1$ appears to form a naked singularity, the formation of naked singularities from smooth initial data of this type is still an open question in mathematical relativity~\cite{Christodoulou:1986zr, Cicortas:2024hpk}. 

We evolve numerically the equations~\eqref{eq:ekg} using the open-source numerical relativity code adapted to spherical polar coordinates \texttt{enGRenage} \cite{Clough:2024yzq}, by the GRTL Collaboration (\href{www.grtlcollaboration.org}{\url{grtlcollaboration.org}})\footnote{We specially thank Katy Clough for developing \texttt{enGRenage} and making it open-source.}. The code uses the BSSN formulation~\cite{Shibata:1995we,Baumgarte:1998te,Nakamura:1987zz,Baumgarte:2012xy}, where the metric is decomposed according to \cite{Arnowitt:1962hi}
\begin{equation}\label{eq:gauge}
    ds^2 = -(\alpha^2 - \beta^r\beta_r) dt^2 +2 \beta_rdtdr+ e^{4\phi}\bar{\gamma}_{ij}dx^i dx^j \, , 
\end{equation}
and the conformally rescaled metric is decomposed in a flat metric, and a deviation from flatness
\begin{equation}
    \bar{\gamma}_{ij} = \hat{\gamma}_{ij} + \epsilon_{ij} \, .
\end{equation}
Above, $\hat{\gamma}_{ij} = dr^2 + r^2(d\theta^2 + \sin^2\theta d\phi^2)$ is the reference flat space metric, and 
\begin{equation}
    \epsilon_{ij} = \mathrm{diag}\Bigl(h_{rr}, r^2h_{\theta\theta}, r^2\sin^2\theta h_{\phi\phi}\Bigr) \, , 
\end{equation}
is the deviation term. We refer the reader to Refs. \cite{Baumgarte:2012xy,Ruchlin:2017com} for more details.

The spacetime dynamics is sourced by the stress-energy of a spherically symmetric, massless scalar field $\Phi$, with stress-energy tensor given in eq.~\eqref{eq:ekg}. We initialize the scalar field with a static, quasi-monochromatic profile 
as in \eqref{eq:ics}, parametrized by an amplitude $\epsilon$, a driving frequency $\omega_0$ (which sets the units in the following), and a pulse location and width $r_0,\sigma$. For all our simulations, we fix
\begin{equation}
    \omega_0=1 \, , \quad r_0 = 150 \, ,\quad \sigma = 20 \, .
\end{equation}
The initial data is static, so the momentum constraint is trivially satisfied. We solve for the Hamiltonian constraint to provide initial data for the conformal factor $\phi$ via
\begin{align}
    \mathcal{H}&\equiv D^2\psi + 2\pi \psi^5\rho\,, \\
    &= \partial_r^2\psi + \frac{2}{r}\partial_r\psi + 2\pi \psi^5\rho=0\,,
\end{align}
where $\psi^4=e^{4\phi}$. The density is defined from contractions of the stress-energy tensor, which at $t=0$ is simply given by the gradients of the scalar field
\begin{align}
    \rho\equiv n^an^bT_{ab}
    &=\frac{1}{2\psi^4}(\partial_r\Phi)^2 \,.
\end{align}
We evolve the BSSN formulation of the equations of motion with the 1+log slicing \cite{Bona:1994dr,Baker:2005vv,Campanelli:2005dd,vanMeter:2006vi} and the Gamma-driver condition \cite{Alcubierre:2002kk}, where the evolution of the lapse and shift is determined by
\begin{align}
    \partial_t\alpha &= -2\alpha K\,,\\
    \partial_t\beta^r &= B^r\,,\\
    \partial_t B^r &= \frac{3}{4}\partial_t\Lambda^r - \eta B^r\,.
\end{align}
Here $\Lambda^r$ is the connection vector \cite{Baumgarte:2012xy} and $\eta$ is a damping constant of order $\sim 1/M_\mathrm{ADM}$. Our simulations are performed in a computational domain with $r\in (0,400]$, evolved until $t=400$, with spatial step-size $\Delta r = 0.05$. We validate the code and demonstrate convergence in Appendix~\ref{sec:appendix}.

When $\epsilon > 1$, the scalar field collapses and forms a BH. Therefore, we search for apparent horizons during the evolution, i.e.\ the outermost marginally outer-trapped surfaces, on which the expansion of outgoing null geodesics $\Theta_+$ vanishes \cite{Thornburg:2003sf},
\begin{equation}
    \Theta_+ \equiv D_i s^i + K_{ij}s^i s^j - K = 0\,,
\end{equation}
where $s^i$ is the unit outward-pointing spacelike normal to the surface. In spherical symmetry \cite{Ruchlin:2017com}, this expression simplifies to
\begin{equation}
    \Theta_+(r) = \frac{4\bar\gamma_{\theta\theta}\,\partial_r\phi + \partial_r\bar\gamma_{\theta\theta}}{e^{2\phi}\,\bar\gamma_{\theta\theta}\sqrt{\bar\gamma_{rr}}} - 2\frac{\bar K_{\theta\theta}}{\bar\gamma_{\theta\theta}}\,,
\end{equation}
and we locate the horizon radius $r_\mathrm{BH}$ by solving $r\,\Theta_+ = 0$ via Newton-Raphson. We then compute the horizon area from the induced metric $q_{AB}$ on the surface $r = r_\mathrm{BH}$ and infer the black hole mass,
\begin{align}
    A_\mathrm{BH} &= \oint \sqrt{\det q}\; d\theta\, d\phi
      = 4\pi\, e^{4\phi}\,\bar\gamma_{\theta\theta}\,\Big|_{r_\mathrm{BH}}\,, \\
    M_\mathrm{BH} &= \sqrt{\frac{A_\mathrm{BH}}{16\pi}} = \frac{1}{2}\, e^{2\phi}\sqrt{\bar\gamma_{\theta\theta}}\,\big|_{r_\mathrm{BH}}\,,
\end{align}
i.e., half the areal radius of the horizon.

\section{Perturbative Regimes}\label{sec:perturbation}

The problem of scalar collapse admits at least two distinct perturbative regimes: for weak amplitudes $\epsilon \ll 1$, where the spacetime is approximately flat at all times; and for large amplitudes $\epsilon > 1$, and at sufficiently late times after BH formation, $t > t_{\rm BH}$. One could also consider perturbations around a critical, discretely self-similar solution, as an expansion in $|\epsilon - 1|\ll 1$, see~\cite{Gundlach:1995kd, Gundlach:2025yje}. In the first case, we consider a perturbative expansion around a Minkowski background 
\begin{equation}
    g = \eta + \epsilon^2 h^{(2)} + \dots \, , \qquad \Phi = \epsilon \Phi^{(1)} + \epsilon^3 \Phi^{(3)} + \dots
\end{equation}
We carry the expansion to next-to-next-to-leading order, $\mathscr{O}(\epsilon^3)$, which is analytically tractable thanks to the spherical symmetry of the problem. 

In the collapsing (or supercritical) regime, and after the BH has formed, the problem reduces to the scattering of a spherically symmetric scalar field off a Schwarzschild BH, 
\begin{equation}
    g = g^{\rm Sch} + \dots \, , \qquad \Phi = \epsilon \Phi^{(1)} + \dots 
\end{equation}
The above set-up would predict that the late-time relaxation of the scalar field is governed by the ringdown of the Schwarzschild geometry. In the following, we review in more detail these two regimes. 

\subsection{Dispersive Regime}

In the sub-critical regime, $\epsilon \ll 1$, the scalar field propagates in an approximately flat spacetime. We can carry out a perturbative expansion tracking the first nonlinear corrections to the scalar field propagation. Let us write the metric in the area gauge, 
\begin{equation}
    ds^2 = -\alpha^2(t_A,r_A)dt_A^2 + a^2(t_A,r_A)dr_A^2 + r_A^2d\Omega^2 \, .
\end{equation}
We introduced an $A$ subindex to indicate that these are area gauge coordinates, as opposed to the numerical coordinates $(t,r)$~\eqref{eq:gauge}. For simplicity, we will work now in the frequency domain. Hence, we write
\begin{equation}
    \Phi = \epsilon e^{-i\omega_0 t_A}\phi_1 + \epsilon^3 \Bigl(e^{-i\omega_0 t_A}\phi_{3,1}+e^{-3i\omega_0 t_A}\phi_{3,3}\Bigr) + \mathrm{c.c.} + \dots 
\end{equation}
The fact that the next-to-leading order piece must take that form is clear from the perturbative structure, and confirmed by numerical evolution in the time-domain~\cite{Cardoso:2026llh}. Given this expansion, we write for the metric components
\begin{equation}
    \begin{aligned}
        \alpha =& 1 + \epsilon^2 (\alpha_{2,0} + e^{-2i\omega_0 t_A}\alpha_{2,2}) + \mathrm{c.c.} + \dots \, , \\ 
        a =& 1+ \epsilon^2 (a_{2,0} + e^{-2i\omega_0 t_A}a_{2,2}) + \mathrm{c.c.} + \dots \, .
    \end{aligned}
\end{equation}
Therefore, when $\epsilon = 0$, the metric is exactly Minkowski, and for finite $\epsilon$ we find the presence of a static perturbation -- due to the mass-energy of the scalar field -- and a time-dependent perturbation\footnote{We emphasize that due to the spherical symmetry of the problem, there is no emission of gravitational waves. This time-dependent piece is not a radiative effect.}. 

To leading order, the scalar field satisfies the wave equation in flat space, so its solutions are of the form
\begin{equation}
    \phi_1 = A \frac{\sin(\omega_0 r_A)}{\omega_0 r_A} \, .
\end{equation}
Einstein equations can be readily solved analytically, to find the metric perturbation:
\begin{equation}
    \begin{aligned}
        a_{20} =& 4\pi A^2\Bigl[1-\frac{\sin^2(\omega_0 r_A)}{\omega_0^2r_A^2}\Bigr] \, , \\
        a_{22} =& \frac{2\pi A^2}{\omega_0^2r_A^2}\sin(\omega_0 r_A) \Bigl[\omega_0 r_A \cos(\omega_0 r_A) - \sin(\omega_0 r_A)\Bigr] \, , \\
        \alpha_{20} =& 4\pi A^2\Bigl[\frac{\sin(2\omega_0 r_A)}{\omega_0 r_A}+2\Bigl(\log r_A-\CI(2\omega_0 r_A)\Bigr)\Bigr] \, , \\
        \alpha_{22} =& \frac{\pi A^2}{2\omega_0 r_A}\sin(2\omega_0 r_A) \, , 
    \end{aligned}
\end{equation}
with $\CI$ the cosine integral function. Remarkably, $\alpha_{20}\sim \log r_A$ as $r_A\to\infty$, signaling an infrared divergence. Indeed, by going to the frequency domain we assume the existence of an infinite, stationary wave-train of scalar field, whose energy formally diverges. This infrared divergence could be cured by introducing a cut-off radius, or rather by working with a wavepacket instead of a monochromatic pulse. Nevertheless, the higher harmonic $\phi_{3,3}$ is not affected by infrared terms. Indeed, the first correction to the Klein-Gordon equation leads to 
\begin{equation}
    \begin{aligned}
        \phi_{33}'' &+ \frac{2}{r_A}\phi'_{33} + 9\omega_0^2\phi_{33} \\
        &=4\omega_0^2 (a_{22}-\alpha_{22})\phi_1+(\alpha_{22}'-a_{22}')\phi_1' \, .
    \end{aligned}
\end{equation}
The solution to this equation is analytic, but we omit it for simplicity. Suffices to note that at large distances $r_A\gg \omega^{-1}_0$, $\phi_{33}$ behaves as
\begin{equation}\label{eq:phi33}
    \phi_{33} \sim - \frac{\pi \cos(3\omega_0 r_A)}{2 \omega_0^2 r_A^2} + \mathscr{O}(r_A^{-3}) \, .
\end{equation}
In particular this means that the higher harmonic satisfies $r \phi_{33} \xrightarrow{r_A\to\infty} 0$, so it does not contribute to a net scalar flux at future null infinity. 

Now, for the nonlinear harmonic oscillating at frequency $\omega_0$, notice that we have 
\begin{equation}\label{eq:phi31_equation}
    \begin{aligned}
        \phi_{31}''&+\frac{2}{r_A}\phi_{31}'+\omega_0^2\phi_{31} = S_1 \, ,  \\
       S_1&= 2\omega_0^2\phi_1(a_{20}-\alpha_{20}) + \phi_1'(\alpha_{20}'+\alpha_{22}'-a_{20}'-a_{22}') \, .
    \end{aligned}
\end{equation}
Notice that the source term contains resonant terms of the form $S_1\supset \sin(\omega_0 r_A),\cos(\omega_0 r_A)$. These resonant pieces may be absorbed by performing a two timescale expansion, in the manner of e.g.~\cite{Balasubramanian:2014cja,Bizon:2015pfa}, where the authors use this expansion to study the backreaction of a massless scalar field in anti-de-Sitter (AdS). There are two main differences in the setup we are interested in: (i) the AdS horizon introduces a cut-off lengthscale that naturally regularizes the infrared singularities present here, manifest e.g. in terms of the form $S_1 \supset \log r_A \sin(\omega_0 r_A)$, and (ii) the radial wave equation in Minkowski with causal boundary conditions has no (quasi)-normal mode solutions, unlike AdS space, which has a complete spectrum of normal modes. However, one can think of the following analysis as stemming from the limit of vanishing cosmological constant of the results of~\cite{Balasubramanian:2014cja, Bizon:2015pfa}. This allows us to absorb the resonant terms as a frequency shift.

Let us modify the original ansatz, to allow the frequency of the leading order piece to drift. We replace $\omega_0 \to \omega_0(1+\epsilon^2 \delta \omega)$. Now eq.~\eqref{eq:phi31_equation} becomes
\begin{equation}
    \phi_{31}''+\frac{2}{r_A}\phi_{31}'+\omega_0^2\phi_{31} = S_1-2 \omega_0^2\delta\omega\phi_1 \, .
\end{equation}
We notice that, at large distances, 
\begin{equation}
    S_1 \sim 8A^2\pi\omega_0^2\Bigl(1-2\log r_A\Bigr)\phi_1 + \mathscr{O}(r_A^{-2}) \, ,
\end{equation}
so we can use the frequency shift $\delta\omega$ to cancel the leading order behavior of $S_1$. We can now safely ignore the unphysical $\log$ term -- any regulator such as a small cosmological constant would eliminate the contribution from this piece. The frequency shift then becomes $\delta\omega = -4\pi A^2$: at leading nonlinear order, the frequency of the pulse will be \emph{redshifted}, and the frequency shift will scale like $\epsilon^2A^2$, consistently with~\cite{Balasubramanian:2014cja,Bizon:2015pfa}. A frequency redshift is precisely what one may expect from the self-gravity of the pulse as it propagates -- our calculation above simply estimates this scaling. 

Note that the analysis carried here is a simplification. Our initial data~\eqref{eq:ics} is only approximately monochromatic, and although $\sigma \gg \omega_0^{-1}$, it is also not stationary. An alternative point of view, based on a time-domain evolution with matching initial data, was carried out in~\cite{Cardoso:2026llh}. We find consistent results whenever the amplitude $\epsilon \ll 1$. 

\subsection{Collapsing Regime}

When $\epsilon > 1$, a BH forms in finite time. The outgoing scalar pulse will feature a transient, which is sensitive to the formation of the horizon, as well as a late time ringdown amenable to a perturbative description. We briefly review here the scattering of scalar waves off a BH of mass $M$.

Let the background geometry be the Schwarzschild spacetime 
\begin{equation}
    ds^2=-fdt_A^2+f^{-1}dr_A^2+r_A^2d\Omega^2 \, , \quad f=1-\frac{2M}{r_A} \, ,
\end{equation}
and let $\Phi = e^{-i\omega t_A} \phi$ be a massless test field with frequency $\omega$. Its propagation is governed by the wave equation 
\begin{equation}\label{eq:rw}
    f \frac{d}{dr_A}\Bigl(f\frac{d\phi}{dr_A}\Bigr) + (\omega^2-V)\phi = 0 \, ,
\end{equation}
with 
\begin{equation}
    V = f\Bigl[\frac{\ell(\ell+1)}{r_A^2}+\frac{2M}{r_A^3}\Bigr] \, .
\end{equation}
At large distances, the field is a superposition of in- and out-going waves
\begin{equation}
    \phi \sim A_{\rm in}(\omega)e^{i\omega r_*} + A_{\rm out}(\omega) e^{-i\omega r_*} \, , \quad r_*\to\infty 
\end{equation}
with $r_*$ the tortoise coordinate. Requiring that the field is ingoing at the horizon constrains the relation between outgoing and ingoing amplitudes. They are related by the BH reflectivity, which is imprinted in the frequency domain waves, see~\cite{Berti:2025hly,Oshita:2023cjz, Okabayashi:2024qbz, Rosato:2024arw, Rosato:2025ulx, Rosato:2026apq}
\begin{equation}\label{eq:refl}
    \calR(\omega) = \Bigl|\frac{A_{\rm out}(\omega)}{A_{\rm in}(\omega)}\Bigr|^2 \, .
\end{equation}
In the time-domain, we can gain intuition from the analytical structure of the Green's function~\cite{Rosato:2025ulx}\footnote{For recent progress in this topic, see~\cite{Arnaudo:2025uos, Arnaudo:2026tcy, DeAmicis:2025xuh, DeAmicis:2026tus, DeAmicis:2026wqd, Rosato:2026moe, Su:2026fvj, Kuntz:2025gdq}.}. We identify two key features: a countably infinite number of poles in the complex plane, corresponding to the \emph{quasinormal mode} frequencies $\omega_n$ for which $A_{\rm in}(\omega_n) = 0$ (equivalently poles of the reflectivity); and a branch cut along the negative imaginary axis, which leads to \emph{tails}. These govern the time-domain response at sufficiently late times, so we can write
\begin{equation}
    \Phi(t_A\gg 0, r_A\to\infty) = \sum_{n=0}^\infty C_n e^{-i\omega_n t_A} + \sum_{p=2\ell+3}^\infty D_p t_A^{-p}  \, .
\end{equation}
We will consider both the frequency- and time-domain responses, and demonstrate that at sufficiently late times, this perturbative description becomes quite accurate. We highlight that there are ample avenues for improvement. First, one could account for the back-reaction of the scalar field on the Schwarzschild BH, carrying the perturbative analysis to higher orders following e.g.~\cite{deCesare:2022aoe, deCesare:2023rmg}. Alternatively one could consider the propagation of a massless scalar field on a Vaidya spacetime, describing a nonrotating BH with a growing mass, following~\cite{Abdalla:2006vb, Redondo-Yuste:2023ipg, Capuano:2024qhv}. 

\section{Results}\label{sec:results}
\begin{figure*}[t!]
    \centering
    \includegraphics[width=\linewidth]{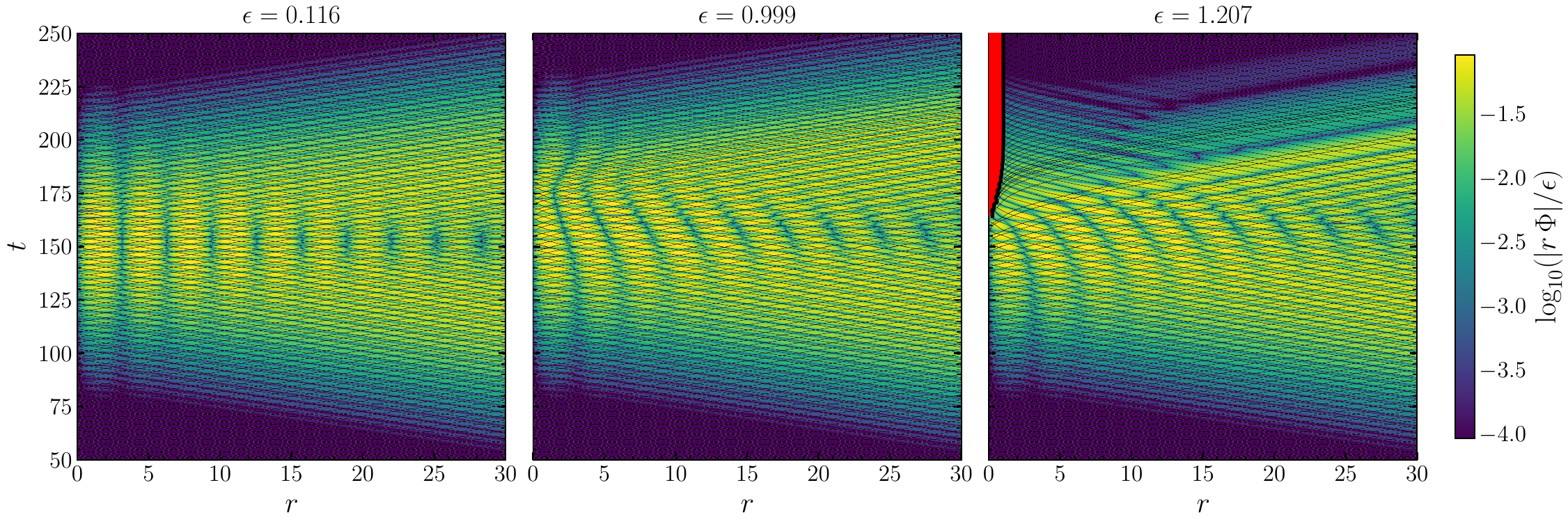}
    \caption{Rescaled profile of the scalar field $\log_{10} |r\Phi|/\epsilon$, as indicated by the colorbar, during the evolution in three different regimes: dispersive $\epsilon = 0.12$ (left), near-critical $\epsilon = 0.999$ (center), and collapsing $\epsilon = 1.21$ (right). Overlaid we show the null rays obtained from integrating~\eqref{eq:null} at each point in the spacetime grid. The null rays accumulate in the strong-field region, as is visible in the near-critical and collapsing simulations. In the collapsing case, we show the area hidden inside the apparent horizon in red. }
    \label{fig:spacetime}
\end{figure*}

We have performed a suite of simulations with increasing amplitude $\epsilon$, keeping $r_0,\sigma$ fixed as indicated in Sec.~\ref{sec:numerics}. In order to obtain the critical amplitude $A_*$, we perform a least-squares fit to
\begin{equation}
    \log M_{\rm BH} = C + \gamma \log(A-A_*) \,, \quad \gamma = 0.37 \, , 
\end{equation}
using Choptuik's critical collapse exponent~\cite{Choptuik:1992jv}. Fitting the BH masses extracted from collapsing simulations we find
$A_* \approx 8.11 \times 10^{-4}$ (for comparison, our last dispersive run has $A = 8.10 \times 10^{-4}$, and our first collapsing run, $A = 8.16 \times 10^{-4}$). We have verified that the value of $A_*$ does not change when working in a different gauge (e.g. setting the shift to zero).

Interpreting results in the strong field region requires a careful understanding of the coordinates, gauge, and slicing. Spherical symmetry is blessed with two privileged directions associated to in- and out-going null rays. Setting $ds^2=0$, null rays satisfy 
\begin{equation}\label{eq:null}
    \Bigl(\frac{dr}{dt}\Bigr)^2 + \frac{2\beta_r}{e^{4\phi}(1+h_{rr})}\frac{dr}{dt} = \frac{\alpha^2-\beta^r\beta_r}{e^{4\phi}(1+h_{rr})} \, ,
\end{equation}
with the $\pm$ sign corresponding to out/ingoing rays respectively. We have traced these rays over our simulations in the different regimes (dispersive, near criticality, and collapsing), and show the resulting causal structure in Fig.~\ref{fig:spacetime}. Overlaid we show the (rescaled) profile of the scalar field. Clearly, null rays accumulate as $\epsilon \to 1$. When $\epsilon > 1$, this accumulation of null rays ultimately leads to the formation of trapped surfaces, shaded in red in Fig.~\ref{fig:spacetime}.

\begin{figure}[t!]
    \centering
    \includegraphics[width=\columnwidth]{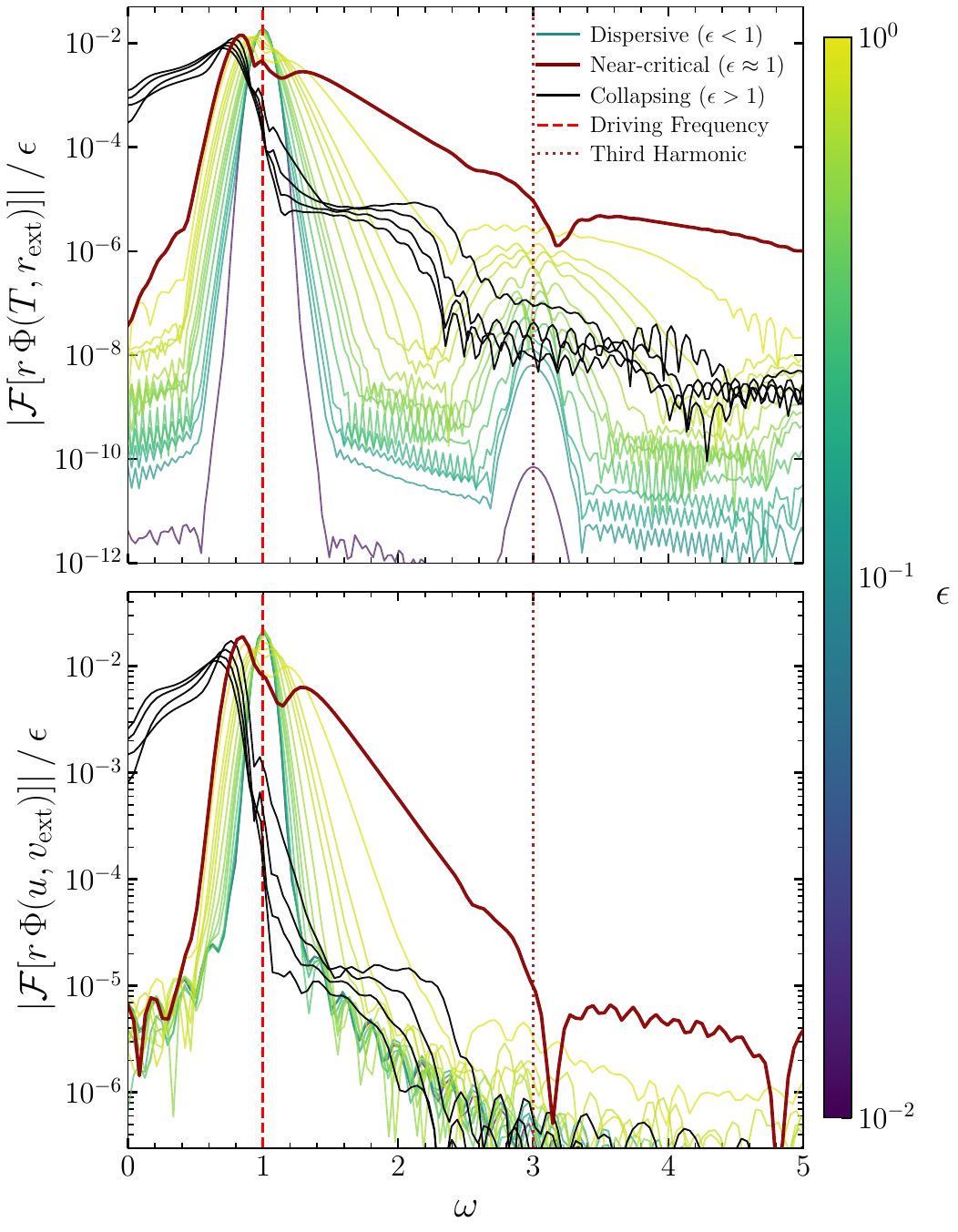}
    \caption{Spectrum of the scalar field $\Phi$ extracted at a timelike $r_{\rm ext}=150$ (top), and at a null slice $v=400$ (bottom), where the Fourier transform is taken with respect to the proper time $T$ and null time $u$, respectively. The colored lines correspond to subcritical evolutions, whose amplitude is indicated by the colorbar. The smaller amplitudes only show peaks at the driving harmonic $\omega=\omega_0=1$ and the first nonlinear harmonic $\omega=3\omega_0=3$ (dotted vertical lines). On the other hand, the near-critical solution, highlighted in red, shows a continuous, flattened spectrum. The black lines correspond to simulations that form BHs, where most of the high frequency content is absorbed by the BH horizon. }
    \label{fig:spectrum}
\end{figure}

The first diagnostic we track is the Fourier spectrum of the field along a given hypersurface. We consider two cases of interest: (i) along a $r=r_{\rm ext}$ surface, where $r_{\rm ext}\gg 1$ so $r\approx r_A$, as a function of the proper time $T$ defined via 
\begin{equation}
    T(t,r_{\rm ext}) = \int_0^t \sqrt{\alpha^2 - \beta_r\beta^r} dt' \, ,
\end{equation}
and (ii) along an ingoing null surface $v=v_{\rm ext}$, as a function of the outgoing null coordinate $u$, approximating the extraction at future null infinity $\mathscr{I}^+$. Our results are shown in Fig.~\ref{fig:spectrum}. Notice that before computing the spectrum we select a window of time ($T$ or $u$) that smoothly isolates the outgoing component of the field -- i.e., the pulse that has bounced from $r=0$, rather than the part of the initial data that propagates outwards. 

A number of features are apparent in Fig.~\ref{fig:spectrum}. First, the spectral amplitude peaks at the driving frequency $\omega_0$, except for the simulations that form BHs. Second, there are clear signs of the excitation of a nonlinear harmonic at $\omega=3\omega_0$. We also see evidence of nonlinear harmonics at $\omega=5\omega_0,7\omega_0$, but, being higher frequency, their amplitudes cannot be extracted accurately enough at our current resolution. These nonlinear harmonics are only visible when extracting along a spacelike instead of a null surface -- the amplitude of the higher harmonics is buried below the noise floor due to numerical error during ray-tracing. The amplitude of these higher harmonics, as seen in the top panel, grows as $\epsilon \to 1$. Third, we also observe a redshift of the peak frequency as $\epsilon \to 1$, in agreement with our perturbative analysis in Sec.~\ref{sec:perturbation}. Finally, we highlight that one can distinguish three qualitatively different regimes: (i) a weakly nonlinear regime, where the spectrum is discrete, and dominated by the driven harmonic and its higher harmonics; (ii) a strongly nonlinear regime (dark red line in Fig.~\ref{fig:spectrum}), very close to the critical threshold, where the spectrum becomes continuous and flattens; and (iii) a qualitatively different behavior once $\epsilon > 1$, which is filtered by the BH absorption of high frequency waves. In the following, we will examine all of these features in more detail.

\subsection{Dispersive Regime}

\begin{figure}
    \centering
    \includegraphics[width=\columnwidth]{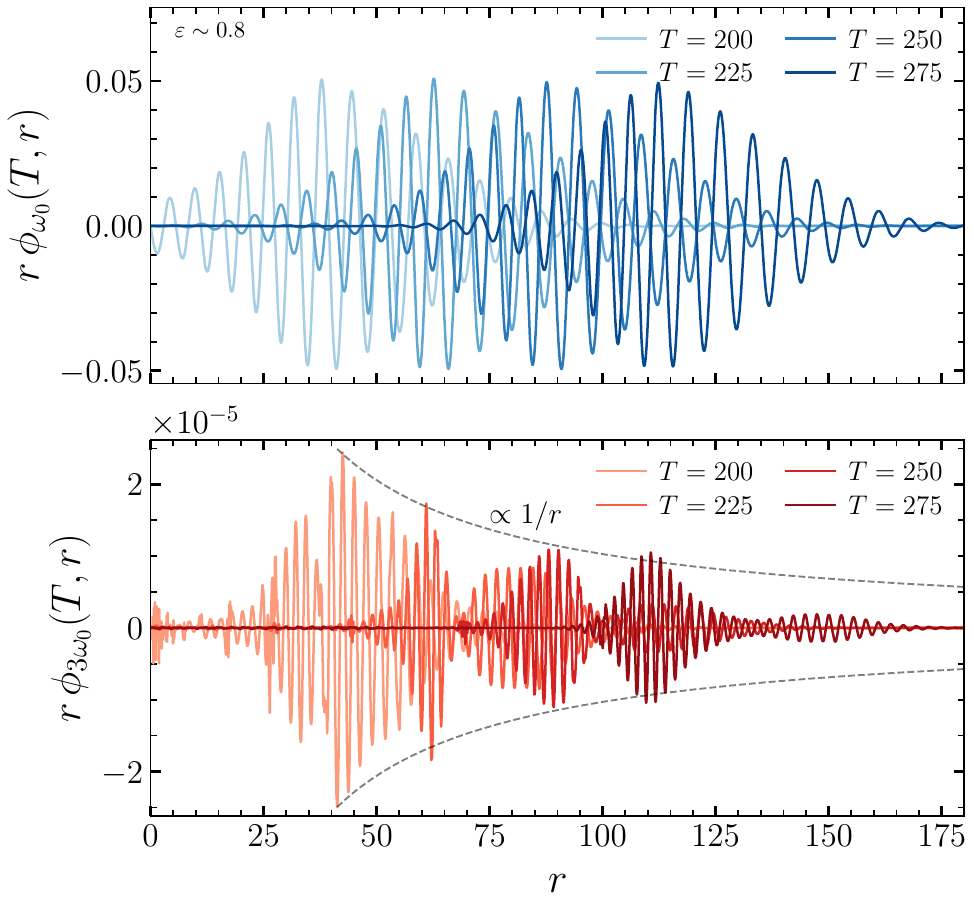}
    \caption{Snapshots of the frequency-isolated radial profiles~\eqref{eq:w_component} of the driven harmonic $\phi_{\omega_0}$ (top), and the higher harmonic $\phi_{3\omega_0}$, for a moderate amplitude simulation, with $\epsilon \approx 0.8$, leading to a maximum luminosity $\mathcal{L}_{\rm peak} \approx 0.03\mathcal{L}_{P}$. The different colors correspond to different times, as indicated by the legend. We rescale the field by $r$ to isolate the leading order peeling contribution. Whereas the driven harmonic propagates without changing amplitude, the nonlinear harmonic decays more rapidly, $r \phi_{3\omega_0} \to 0$, as anticipated in~\cite{Cardoso:2026llh}. As a black envelope we show the scaling predicted perturbatively (in the case of an infinite train), which only provides a qualitative agreement.}
    \label{fig:propagation}
\end{figure}

Let us begin by analyzing the amplitude of the leading harmonics present in the signal: the driven harmonic with frequency $\omega_0$, and the first nonlinear harmonic, with frequency $3\omega_0$. We extract these two harmonics sufficiently far away from the strong-field region, to ensure we are not sensitive to gauge dynamics. Next, we isolate each frequency component by defining
\begin{equation}\label{eq:w_component}
    \phi_\omega(T,r_{\rm ext}) = \mathcal{F}^{-1}\Bigl[W_\omega  \mathcal{F}[\Phi(T, r_{\rm ext})]\Bigr] \, , 
\end{equation}
where $\mathcal{F}$ denotes the (temporal) Fourier transform with respect to proper time $T$, and $W_\omega$ is a Hann window function with support in $\omega\pm 0.5$. This allows us to reconstruct the evolution and propagation of each frequency component independently\footnote{We cannot cleanly do the same separation of frequency components when extracting the field along a $v=\mathrm{const.}$ surface due to noise stemming from numerically finding the null rays.}. We show their evolution for a moderate amplitude in Fig.~\ref{fig:propagation}. The top (respectively, bottom) panel shows the evolution of $\phi_{\omega_0}$ (respectively, $\phi_{3\omega_0}$), once the pulse has ``bounced'' off the origin. The scattered component of the pulse with frequency $\omega$ decays as $1/r$ -- therefore, the envelope of the plotted quantity $r \phi_{\omega_0}$ is constant. On the other hand, the higher harmonic has a faster spatial decay, qualitatively consistent with the perturbative calculation, $\phi_{3\omega_0} \sim r^{-2}$, c.f.~\eqref{eq:phi33}. Its shape is also distorted, although some contamination from noise in the Fourier transform is evident: as the nonlinear harmonic propagates its amplitude decays, and at some radius it drops below the spectral noise floor.

\begin{figure*}[t]
    \centering
    \includegraphics[width=\linewidth]{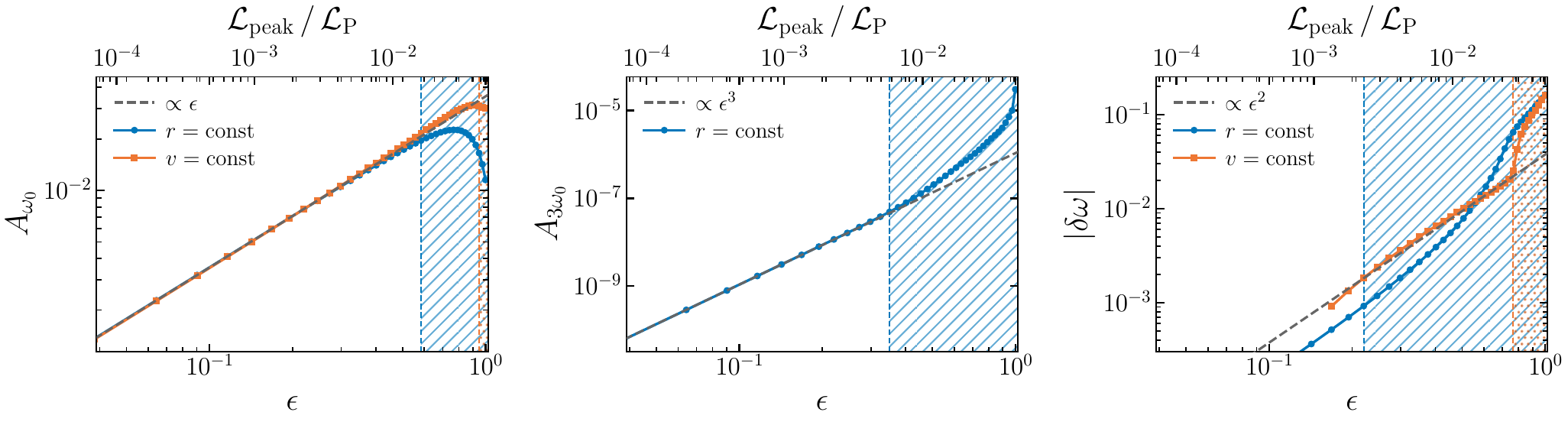}
    \caption{Nonlinear diagnostics as a function of the normalized amplitude of the initial data $\epsilon$ (bottom axis), and the maximum luminosity in units of Planck's luminosity (top axis), compared to their predicted scaling within perturbation theory. \textbf{Left:} Amplitude of the driven harmonic $\omega=\omega_0=1$ extracted along a $r=150=\mathrm{const.}$ slice, and measured with respect to proper time $T$ (blue), and along an ingoing null slice $v=400=\mathrm{const.}$ (orange). The dashed lines show the fit of the first few points to a linear scaling in the amplitude, $\epsilon$. \textbf{Center:} Same but for the third harmonic $\omega=3\omega_0$, which is buried behind the numerical noise for the null slice. Here the dashed line shows the expected $\epsilon^3$ scaling. \textbf{Right:} Same but for the frequency shift $\delta\omega = \omega_{\rm peak}-\omega_0$, which perturbatively we expect to scale as $\epsilon^2$. In all panels, the vertical dashed lines and shaded regions mark the values of $\epsilon$ for which the deviations with respect to the perturbative scaling are larger than $5\%$. We can understand these ranges as the values of $\epsilon$ beyond which perturbation theory around flat space, up to $\mathscr{O}(\epsilon^3)$, breaks down.}
    \label{fig:diagnostics}
\end{figure*}

We now turn to investigate three diagnostics of nonlinear dynamics: the amplitude of the driving harmonic $\omega_0$, the nonlinear harmonic $3\omega_0$, and the frequency shift of the driven harmonic. The amplitude of each of the harmonics is extracted directly from the Fourier transform, see Fig.~\ref{fig:spectrum}. As for the frequency shift, we simply find the frequency at which the maximum spectral amplitude is achieved $\omega_{\rm peak}$, and define $\delta\omega=\omega_{\rm peak}-\omega_0$. All of these quantities extracted at both timelike and null slices are shown in Fig.~\ref{fig:diagnostics}. Based on the perturbative analysis of Sec.~\ref{sec:perturbation}, we expect the driven harmonic with frequency $\omega_0$ to scale as $A_{\omega_0} \sim \epsilon$; the higher harmonic with frequency $3\omega_0$ to scale as $A_{3\omega_0} \sim \epsilon^3$, and the frequency shift to scale as the slow time, $\delta \omega \sim \epsilon^2$. These scalings accurately hold for $\epsilon \ll 1$ in Fig.~\ref{fig:diagnostics}. For all except for the third harmonic (which lies below the noise floor) we show the results obtained both at constant $r$ and constant $v$ slices.

In all three cases, we observe deviations from these scalings before $\epsilon = 1$. We define a proxy for ``breakdown'' of the perturbative description to this order as the value of $\epsilon$ for which a given diagnostic deviates from the perturbative prediction by more than $5\%$. This threshold is arbitrary -- nevertheless, it provides some qualitative guidance. Similarly, the exact value of $\epsilon$ at which deviations are first observed is slightly different for all three diagnostic quantities, and the deviations appear later whenever we extract along an ingoing null slice. One reason is that some of the nonlinear features, e.g., the higher harmonic, decay fast as they propagate. We have indeed verified that extracting at larger values of $r$ moves the ``breakdown'' value of $\epsilon$ closer to the critical threshold $\epsilon = 1$. This means that an observer closer to the strong-field region may measure strong nonlinear effects, that an asymptotic observer in the far field region remains oblivious to, and cautions us against interpreting the linearity of physical processes looking only at asymptotic quantities. Nevertheless, we consistently find deviations -- even for asymptotic observers close to future null infinity (see orange line in Fig.~\ref{fig:diagnostics}) with respect to the perturbative trend at some $\epsilon < 1$, i.e., perturbation theory breaks down before BH formation.

It is illustrative to understand this threshold not in terms of $\epsilon$ but rather in terms of a physical quantity. A natural choice is the maximum luminosity, in natural units, $\mathcal{L}_{\rm peak} / \mathcal{L}_P$. The details on how the maximum luminosity is measured are given in Appendix~\ref{app:luminosity}, and follow closely~\cite{Cardoso:2018nkg}. When $\epsilon \ll 1$, we have that $\mathcal{L}_{\rm peak} \propto \epsilon^2$, but this scaling breaks down as $\epsilon\to1$. Fig.~\ref{fig:diagnostics} shows evidence that the breakdown of the perturbative description occurs when $\mathcal{L}_{\rm peak}/\mathcal{L}_P \gtrsim 10^{-2}$. This motivates considering the maximum luminosity as a natural scale that quantifies how ``nonlinear'' a process can be, or even as a perturbative parameter quantifying the strength of nonlinear effects. 

So far, we conclude that dispersive solutions cover two qualitatively distinct regimes: when $\epsilon$ is sufficiently small, the spectrum (Fig.~\ref{fig:spectrum}) is discrete, and the quantitative predictions of perturbation theory up to $\mathscr{O}(\epsilon^3)$ are accurate. We encounter a distinct regime when $\epsilon \lesssim 1$. The spectrum becomes continuous, and flattens; and the quantitative scalings of perturbation theory break down -- e.g., the amplitude of the higher harmonic $\omega=3\omega_0$ is \emph{larger} than what perturbation theory would predict. 

\subsection{Collapsing Regime}

Now we focus on the simulations that form a BH, i.e., with $\epsilon > 1$. In that case, we distinguish two stages in the outgoing pulse: a first stage, which is driven by the initial data; and a final stage, after a horizon forms, that is governed by the quasinormal modes of the newly formed BH~\cite{Purrer:2004nq}. 

\begin{figure}
    \centering
    \includegraphics[width=\columnwidth]{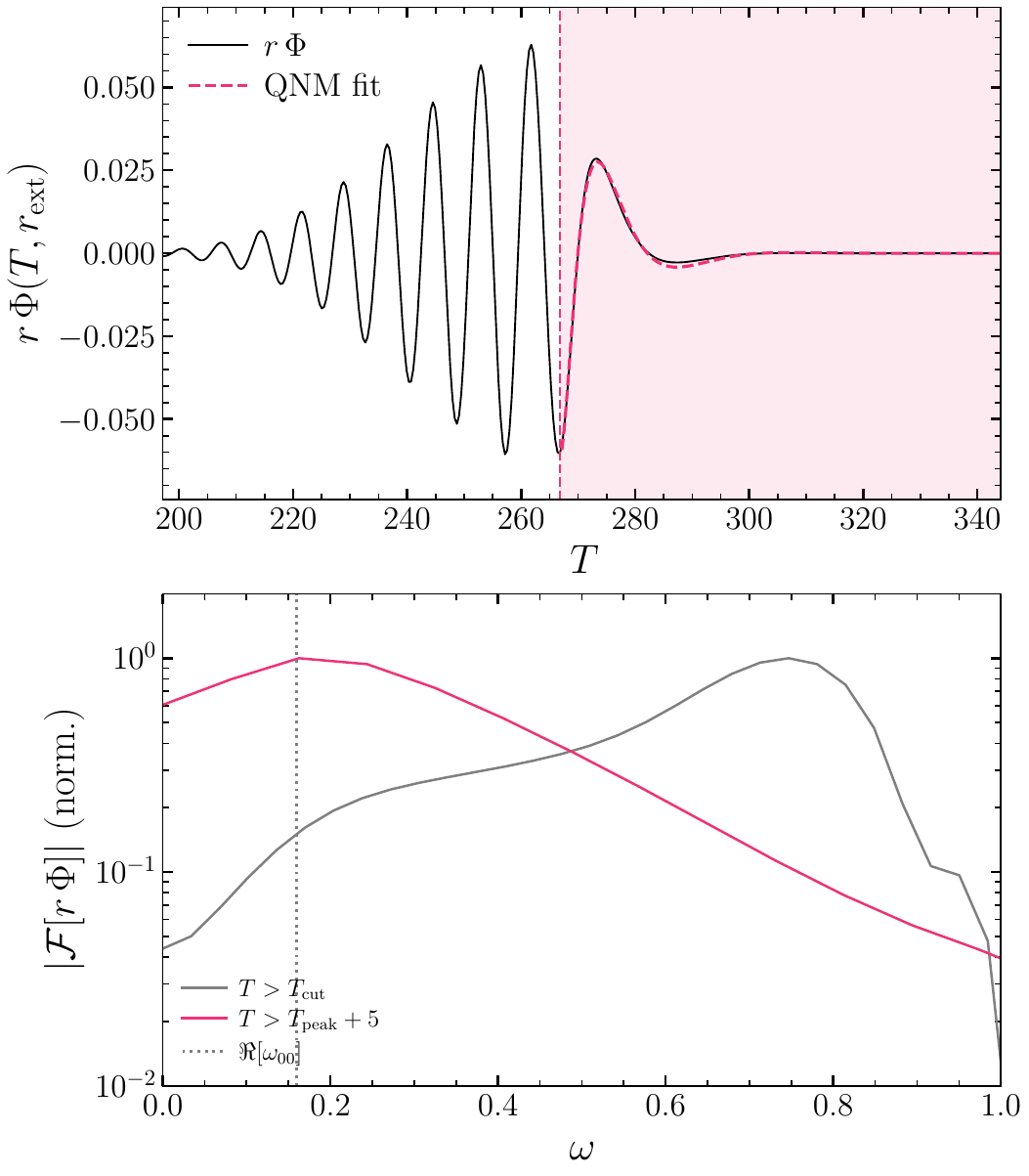}
    \caption{\textbf{Top:} Scattered pulse from a supercritical evolution $\epsilon = 1.1$ extracted at $r_{\rm ext}=100$. The red dashed line shows the fit of the late-time signal to a superposition of the scalar fundamental mode and the first overtone of a Schwarzschild BH. \textbf{Bottom:} Fourier transform of the full signal in the top panel (in black), and of the signal only after $T > T_{\rm peak}+5\sim 265$ (red). Recall that $T$ is the proper time experienced by an observer located at constant $r$. For reference, we include the BH reflectivity~\eqref{eq:refl} (blue), and the fundamental scalar spherically symmetric QNM frequency (vertical dashed line). }
    \label{fig:collapsing}
\end{figure}

We analyze the outgoing pulse in a collapsing evolution in Fig.~\ref{fig:collapsing}. We extract the pulse at a fixed radius $r_{\rm ext}=100$, sufficiently far away from the strong-field region. These two distinct regimes are clearly visible in the top panel. At late times, the scalar field decay can be approximated by a superposition of quasinormal modes. In particular, we construct a model containing the fundamental mode and first overtone, 
\begin{equation}
    r\Phi \approx \sum_{n=0}^1 A_n \cos(f_n T + \phi_n)e^{-T/\tau_n} \, , 
\end{equation}
where $f_0 M_{\rm BH}= 0.1105$, $\tau_0 /M_{\rm BH} = 9.5333$, $f_1 M_{\rm BH} = 0.0861$, and $\tau_1/M_{\rm BH} = 2.8731$ are the frequencies and the damping times of the spherically symmetric, scalar quasinormal modes of a nonrotating BH~\cite{Berti:2009kk}. Fig.~\ref{fig:collapsing} demonstrates that the signal at late times is well approximated by this superposition of quasinormal modes. Note that the fit could possibly be improved by adding additional overtones~\cite{Berti:2025hly}.

We also investigate the signal in the frequency domain, in the bottom panel. The black and red lines correspond to the spectral amplitude from $T>T_{\rm cut}\sim 160$, and from $T>T_{\rm peak}+5$, respectively. When including the whole signal, the spectrum is peaked near the (redshifted) driving frequency $\omega \lesssim 1$. However, when cutting\footnote{To avoid spectral leakage issues we include a Tukey taper window with $\alpha=0.1$.} the signal to include only the ``ringdown'' stage, the resulting spectrum is peaked around the real part of the dominant scalar quasinormal mode of the resulting BH. Consistently with e.g.~\cite{Rosato:2025ulx} we find that at high frequencies the signal does not decay exactly like the BH reflectivity, but rather the spectral amplitude carries imprints of the initial data. 

\subsection{Near-Critical Regime }

\begin{figure*}[t!]
    \centering
    \includegraphics[width=\linewidth]{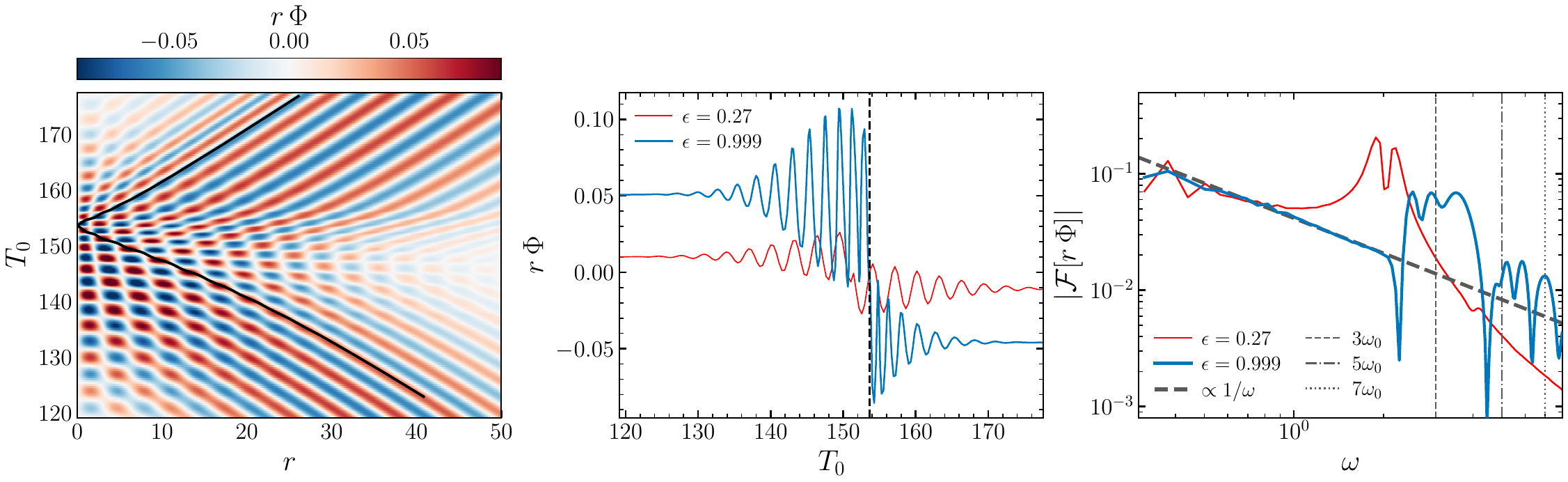}
    \caption{Scalar field profile and spectrum along a ``bouncing'' null geodesic for the near-critical run with $\epsilon = 0.999$. \textbf{Left:} Spacetime profile of the scalar field, and the bouncing null geodesic along which we extract, in black. \textbf{Center:} Profile of the field along such a geodesic, parametrized by the proper time at the origin $T_0$, for two simulations with different values of $\epsilon$, as indicated in the legend. \textbf{Right:} The Fourier transform of the central panel. The near-critical simulation shows a clear $1/\omega$ scaling at low frequencies, in agreement with~\eqref{eq:prediction}. In addition, peaks at the frequencies of the higher harmonics $3,5,7\omega_0$ are visible, and indicated with dashed lines.}
    \label{fig:near_critical}
\end{figure*}

Finally we examine in more detail the dynamics close to criticality, i.e., for $\epsilon \approx 1$. Since we need to keep sufficient resolution in a large computational domain, we do not go beyond $|\epsilon - 1| \sim 10^{-3}$. Pushing beyond this limit would be a natural direction of continued research. 

Let us analyze the spectrum in the limit $\epsilon \to 1$. Fig.~\ref{fig:spectrum} shows a clear flattening of the spectral amplitude and a transition from a discrete to a continuous spectrum. For instance, the frequency bin $\omega=2\omega_0$ has a negligible spectral amplitude as $\epsilon \to 0$, but not as $\epsilon \to 1$. However one must keep in mind that Fig.~\ref{fig:spectrum} discusses the outgoing pulse as seen by observers (timelike or null) in the asymptotic region. It would be illustrative to consider instead an observer that crosses the strong-field region $r\sim 0$ when the spacetime is most dynamical.

We construct a null geodesic $\gamma_{0}: (t_0(\lambda),r_0(\lambda))$ such that it bounces off the origin $r_0(\lambda_*)=0$ when the lapse at the origin is minimized. The geodesic $\gamma_0$ is ingoing for $\lambda<\lambda_*$ and outgoing for $\lambda>\lambda_*$. This geodesic is shown in the left panel of Fig.~\ref{fig:near_critical} (black line). The field along that ray is shown in the central panel of Fig.~\ref{fig:near_critical}. As expected, at early and late times, $r\Phi$ is constant along the null ray. If $\epsilon \to 0$, this constant would be exactly the same at all times. However in the presence of nonlinear dynamics, this is no longer the case, and we observe both rapid oscillations and a mismatch between $r\Phi$ along the in and out-going null rays. The right panel of Fig.~\ref{fig:near_critical} shows the spectral amplitude seen by such an observer, with respect to the proper time at the origin $T_0$. 

Now recall that near-critical solutions have been shown to be approximately self-similar~\cite{Choptuik:1992jv}. In particular, for the massless scalar field considered in this work, the approximate discrete self-similarity (DSS) for $X = r\Phi$ takes the form 
\begin{equation}
    X(\tau+\Delta) \approx X(\tau) \, , \quad \tau = -\log|T_0^*-T_0| + \kappa \, ,
\end{equation}
with $\Delta\approx 3.44$ the period characterising DSS, and where $T_0$ is the proper time for an observer at the origin. Above, $T_0^*$ is the proper time where a singularity first forms. This approximate DSS can be exploited to make a very concrete prediction about the spectral shape of $X$, when taking the Fourier transform conjugate to the proper time $T_0$. Indeed, a $\tau$-periodic solution can be written as 
\begin{equation}
    X = \sum_{n\in\mathbb{Z}}X_n e^{i\omega_n \tau} \, , \qquad \omega_n = \frac{2\pi n}{\Delta} \, ,
\end{equation}
for some coefficients $X_n$. A straightforward calculation shows that 
\begin{equation}\label{eq:prediction}
    \begin{aligned}
        \mathcal{F}_{T_0}[X] &= \sum_nX_ne^{i\omega_n \kappa}\int dT_0 e^{-i\omega T_0}|T_0^*-T_0|^{-i\omega_n}  \\
        &=\sum_n \tilde{X}_n |\omega|^{-1-i\omega_n} \, .
    \end{aligned}
\end{equation}
Therefore the spectral amplitude decays, to leading order, as $1/|\omega|$, with a modulation period $\Delta$. Fig.~\ref{fig:near_critical} shows that we recover exactly this scaling for near-critical simulations (see blue line). For comparison we show the same analysis for a simulation far from criticality (red line), where the spectrum quickly deviates from the $1/\omega$ scaling. The near-critical, blue line, also shows clear hints of higher harmonics as peaks at frequencies $\omega=3\omega_0, 5\omega_0, 7\omega_0$.

Of course, this scaling is only approximate -- DSS itself is only approximate for solutions that are not exactly critical. Nevertheless, we find it remarkable that this is sufficient to explain the flattening of the spectrum close to criticality. An open question, at this stage, is whether perturbation theory around the critical solution can explain other features of the dynamics of the scalar field close to criticality, in particular, the deviations of the scalings of Fig.~\ref{fig:diagnostics}.

\section{Discussion}\label{sec:conclusions}

Perturbation theory is remarkably effective at describing many gravitational processes, such as the two-body dynamics (as an expansion in the velocity, in the gravitational constant, or in the mass ratio), and the black hole ringdown. This is sufficient motivation to investigate regimes where perturbation theory around simple spacetimes loses accuracy, as these are the regimes where gravitational dynamics becomes strongly nonlinear. The collapse of a scalar field, near the critical threshold, is one such regime that we explored in this work.  

We study the collapse of a quasi-monochromatic pulse, driving the spacetime with a wavelength $\omega_0^{-1}$. The amplitude of the initial data $\epsilon = A/A_*$ (normalized with respect to the critical amplitude) emerges as a natural perturbative parameter: $\epsilon \to 0$ corresponds to the propagation of a test scalar field on flat spacetime, and $\epsilon = 1$ corresponds to the critical solution where seemingly a naked singularity develops~\cite{Christodoulou:1986zr, Choptuik:1992jv}. This parameter is directly related to the maximum luminosity in natural units $\mathcal{L}_{\rm peak}/\mathcal{L}_{P}$, which we identify as a physically motivated perturbative scale. 

In the regime where $\epsilon \ll 1$, the scalar field behaves as a test field propagating in flat space. Accounting for the leading order nonlinear effects shows the appearance of a higher harmonic -- a peak in the spectrum at frequency $3\omega_0$, whose amplitude scales as $\epsilon^3$, and a shift in the driving frequency $\omega_0 \to \omega_0 + \epsilon^2 \delta \omega$. We identify both effects in numerical simulations and demonstrate that they obey the perturbative scalings when $\epsilon$ is sufficiently small. In addition, we also show that the higher harmonics do \emph{not} propagate to future null infinity, exhibiting instead a faster radial decay (peeling), confirming the perturbative predictions of~\cite{Cardoso:2026llh}. This is in contrast with higher harmonics in black hole spacetimes, which can indeed reach future null infinity~\cite{Cheung:2022rbm, Mitman:2022qdl}. 

Our work identifies a breakdown of the perturbative description around flat spacetime, at least up to $\mathscr{O}(\epsilon^3)$. In this regime, the amplitude of the excited harmonics stops scaling as a simple power law with $\epsilon$. The spectral amplitude of the outgoing scalar field flattens, meaning that energy is deposited at frequencies \emph{outside} of the natural harmonics of the problem. This is observed both for timelike and null observers, in the far-field region. In particular we identify a power law $\omega^{-1}$ scaling of the spectrum close to criticality, which follows from the approximate discrete self-similarity of the near-critical solutions. This shows that even though the problem appears non-perturbative, as an expansion in $\epsilon \ll 1$, it may become perturbative in $|\epsilon -1|$, as an expansion around a nonlinear, critical solution. We emphasize here that this phenomenon is not strictly related to turbulence, despite appearing as a power-law-like spectral amplitude. An identification of turbulent phenomena would require sufficiently long trapping~\cite{Benomio:2024lev, Redondo-Yuste:2025hlv, Siemonsen:2025fne, Siemonsen:2025ucx, Siemonsen:2025wib}, or a steady driving of the spacetime~\cite{Ma:2025rnv}. It would be interesting to verify that the flattening of the spectrum that we observe (for null observers that get sufficiently close to the strong-field region) is reproduced for other classes of initial data, including in particular axisymmetric critical collapse in vacuum. In light of recent work~\cite{Baumgarte:2023tdh, Baumgarte:2026pqr}, it is likely that this set up exhibits richer -- and non-universal -- spectral features near criticality. 

We have not studied whether resummations of perturbation theory around flat space, or perturbations around the critical solution, can explain the behavior shown in Fig.~\ref{fig:diagnostics}, and the flattening of the spectrum in Fig.~\ref{fig:spectrum}. The power-law scaling of the spectral amplitude near criticality, though, hints at the problem remaining perturbative, albeit with respect to a different background spacetime. An avenue to further investigate this is to consider \emph{extremal} critical collapse~\cite{East:2025nfb, Angelopoulos:2026bez}, where the critical solution is well understood: an extremal BH. Other future research directions include the collapse of axisymmetric gravitational waves~\cite{Baumgarte:2023tdh, Baumgarte:2026pqr}. We conjecture that the maximum luminosity at which backreaction becomes important also lies well below Planck's luminosity (see e.g.~\cite{Flanagan:1993zz,Cardoso:2018nkg}), which would provide further evidence that $\mathcal{L}_{\rm peak}/\mathcal{L}_{P}$ acts as a natural perturbative parameter, even in the absence of symmetries. This could explain why quasicircular, comparable mass BH mergers are well explained by different perturbative approaches e.g. in the ringdown, with nonlinear effects remaining small, as $\mathcal{L}_{\rm peak}/\mathcal{L}_{P} \sim 10^{-3} \ll 1$.  

The ultimate goal is to learn lessons that can be extrapolated to more complex problems, such as binary BH mergers. Recent results~\cite{Zhu:2026mhn} demonstrate that at sufficiently large center of mass energies, the maximum energy emitted deviates from perturbative analysis~\cite{Smarr:1977fy,DEath:1992mef, DEath:1992nmz, DEath:1992plq}. The scattering of pulses of matter also shows comparably high luminosities~\cite{Patino:2026jlq}, appearing as an exciting laboratory to test nonlinear dynamics in General Relativity. In a similar context, other works have explored the validity of perturbative methods (BH perturbation theory and post-Minkowskian expansions) to predict the scattering angle in scattering encounters between two black holes~\cite{Rettegno:2023ghr,Swain:2026dfr, Clark:2026bgg}. Extending the perturbative predictions further into the nonlinear regime, and comparing with nonlinear simulations at increasingly high center-of-mass energies appears as a promising research direction.

\begin{acknowledgments}

We thank Vitor Cardoso, Katy Clough, Pau Figueras, and Christoph Kehle for valuable discussions and comments. We would also like to thank the GRTL collaboration (\url{www.grtlcollaboration.org}) and in particular Katy Clough for their code development work.
J.~R.-Y. is supported by JSPS Postdoctoral Fellowships for Research in Japan, and by NSF Grants No.~AST-2307146, No.~PHY-2513337, No.~PHY-090003, and No.~PHY-20043, by NASA Grant No.~21-ATP21-0010, by John Templeton Foundation Grant No.~62840, by the Simons Foundation [MPS-SIP-00001698, E.B.], by the Simons Foundation International [SFI-MPS-BH-00012593-02], and by Italian Ministry of Foreign Affairs and International Cooperation Grant No.~PGR01167. JCA acknowledges funding from the Department of Physics at MIT through a CTP Postdoctoral Fellowship. 
We thank the Simons Collaboration in Black Holes and Strong Gravity, whose last meeting provided a stimulating environment for discussions that helped shape this work.

This work used Stampede3 at the Texas Advanced Computing Center through allocations PHY090003  and PHY260066 from the Advanced Cyberinfrastructure Coordination Ecosystem: Services \& Support (ACCESS) program \cite{ACCESS}, which is supported by U.S. National Science Foundation grants 2138259, 2138286, 2138307, 2137603, and 2138296. The authors acknowledge the Texas Advanced Computing Center (TACC) at The University of Texas at Austin for providing computational resources that have contributed to the research results reported within this paper. URL: \url{www.tacc.utexas.edu}.

\end{acknowledgments}

\bibliography{biblio}
\clearpage 
\onecolumngrid

\appendix

\section{Code validation and convergence}\label{sec:appendix}

A solid test of the accuracy of our evolution in the most nonlinear regime is provided by the critical scaling of the black hole mass. Choptuik~\cite{Choptuik:1992jv} showed that in this set-up the resulting BH mass scales as $M_{\rm BH} \approx C\,|\epsilon - 1|^\gamma$, with $\gamma \approx 0.37$. We have tested that our simulations recover this scaling within reasonable accuracy. 

\begin{figure}[h!]
    \centering
    \includegraphics[width=0.9\linewidth]{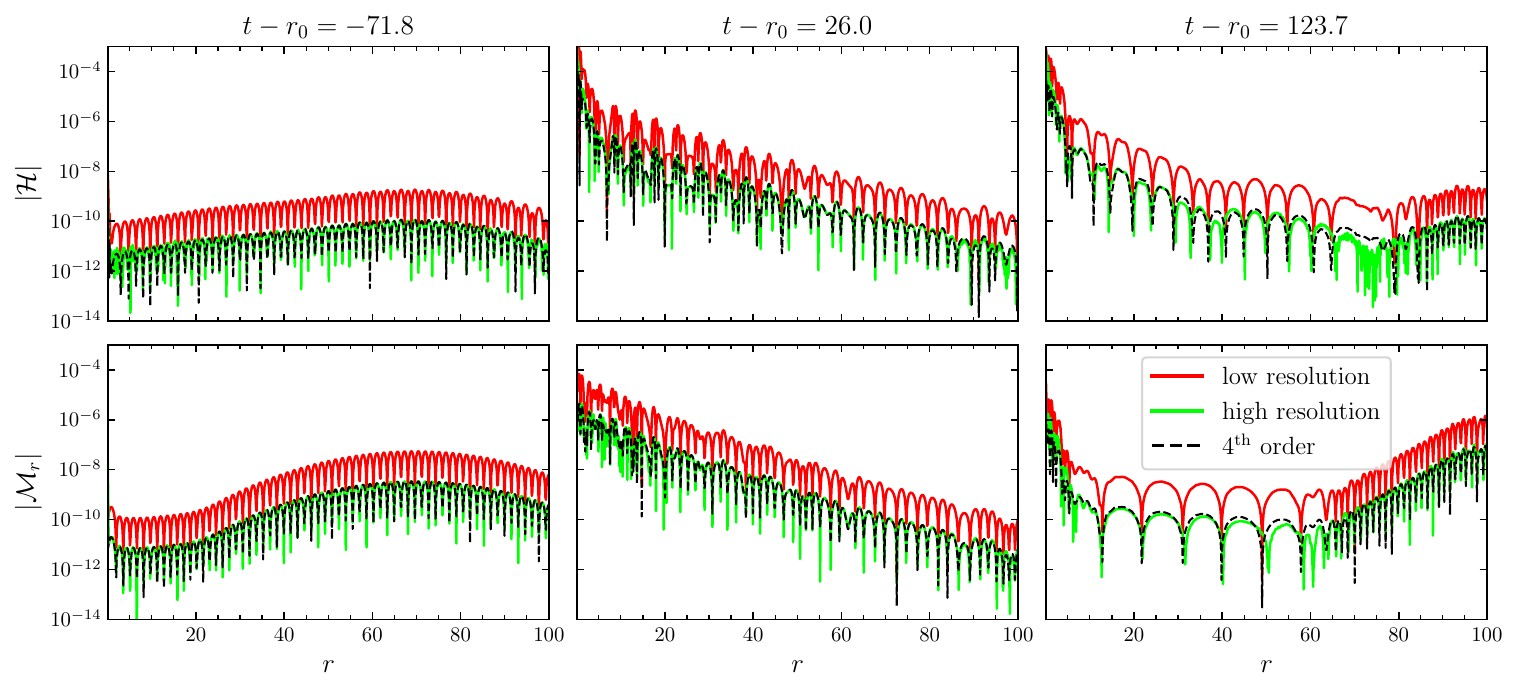}
    \caption{Convergence test of the constraint violations (Eqs. (46-47) in \cite{Ruchlin:2017com}) for a pulse with $\epsilon\approx0.92$ and centred at $r_0=150$, showing radial lineouts of the Hamiltonian $|\mathcal{H}|$ (top row) and momentum $|\mathcal{M}_r|$ (bottom row) constraints at three stages of the evolution. Results are shown for two resolutions, with the low-resolution violation rescaled by $1/2^{4}$ (dashed black line) to demonstrate 4th order convergence, consistent with the used finite-difference stencils.}
    \label{fig:convergence}
\end{figure}

We solve the Hamiltonian constraint to construct the initial data, as the momentum constraint is trivially satisfied for a scalar pulse at rest. We nevertheless monitor both constraint violations throughout the evolution to validate the code. We evolve a pulse with $\epsilon \approx 0.92$ centred at $r_0 = 150$ at two resolutions, $\Delta r_\mathrm{LR} = 0.1$ and $\Delta r_\mathrm{HR} = \Delta r_\mathrm{LR}/2$. Figure~\ref{fig:convergence} shows the constraints converging at fourth order, consistent with our finite-difference stencils.

\section{Determination of the Luminosity}\label{app:luminosity}

\begin{figure}[h!]
    \centering
    \includegraphics[width=\linewidth]{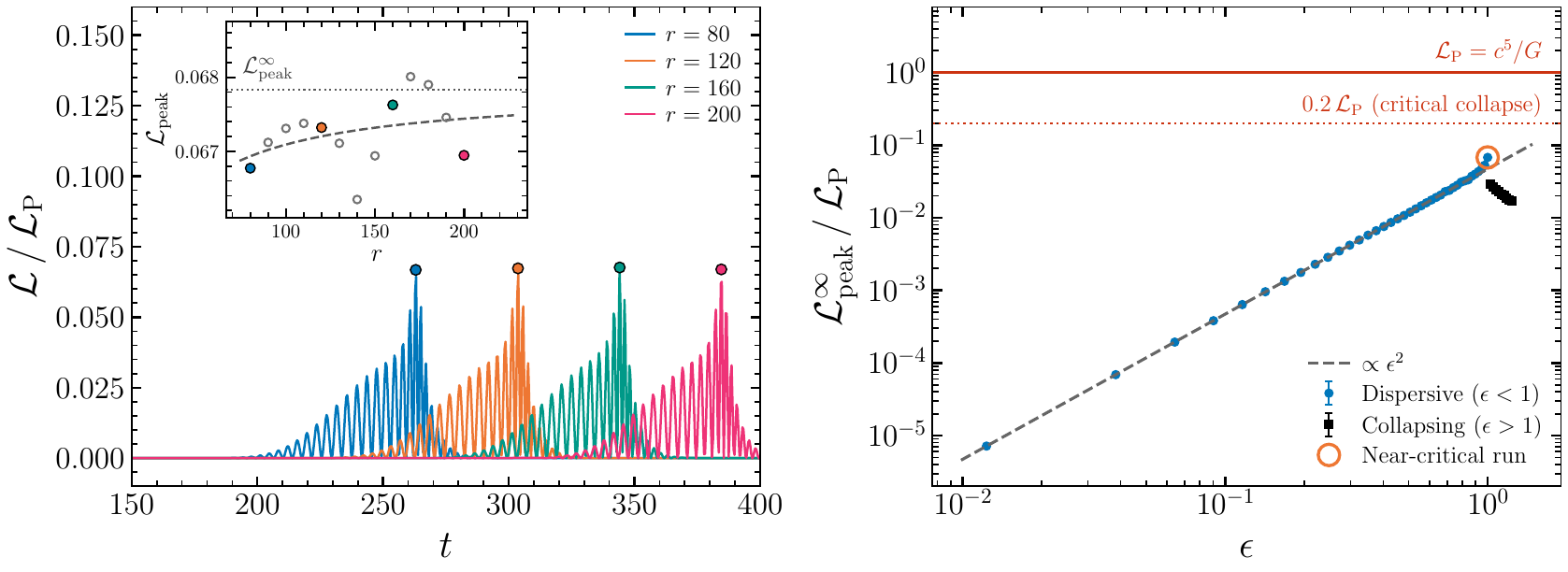}
    \caption{\textbf{Left:} Dimensionless luminosity for the near-critical run with $\epsilon = 0.999$, as a function of time, for four different extraction radii (see legend). From each of these extraction radii, we extract a value of $\mathcal{L}_{\rm peak}$, that we fit following~\cite{Cardoso:2018nkg} to extract the peak luminosity at infinity (see inset). \textbf{Right:} Peak luminosity at infinity in dimensionless units as a function of the initial data amplitude $\epsilon$. The maximum luminosity is achieved close to criticality, but lies well below Planck's luminosity.}
    \label{fig:luminosity}
\end{figure}

In order to determine the maximum luminosity for each evolution, we follow Ref.~\cite{Cardoso:2018nkg}. Here we briefly review this construction. Spherical symmetry admits a natural way of defining a conserved flux 
\begin{equation}
    S^a = K^b T^a_b \, , \quad K^a = \epsilon^{ab} \nabla_b r_A \, , 
\end{equation}
where $K^a$ is the Kodama vector~\cite{Kodama:1979vn}, $T_{ab}$ the stress energy tensor of the scalar field, and recall that $r_A$ is the area radius. It follows that the flux vector $S^a$ is conserved
\begin{equation}
    \partial_a \Bigl(\sqrt{-g}S^a\Bigr)=0 \, .
\end{equation}
The luminosity is simply defined from the radial component of the flux vector,
\begin{equation}
  \mathcal{L}(t, r)
    = -\oint_{S_r} d\theta\, d\varphi \, \sqrt{-g}\; S^r \,,
  \label{eq:lum-def}
\end{equation}
signed so that outgoing radiation has $\mathcal{L} > 0$. The measurement is illustrated in the left panel of Fig.~\ref{fig:luminosity} for the
loudest run of the amplitude scan, at $\epsilon = 0.999$.  We record $\mathcal{L}(t)$ at thirteen extraction radii
$r \in \{80, 90, \ldots, 200\}$ and locate the peak of the outgoing burst.  The residual finite-radius error is then removed by fitting the radial decay of the luminosity, following~\cite{Cardoso:2018nkg}. We also account for the intrinsic oscillations of the flux (see inset panel of Fig.~\ref{fig:luminosity}).
\begin{equation}
  \mathcal{L}_{\rm peak}(r)
    = \mathcal{L}_{\rm peak}^{\infty} + \frac{c_1}{r} + \frac{c_2}{r^2} \,.
  \label{eq:lum-extrap}
\end{equation}
The variation of
$\mathcal{L}_{\rm peak}(r)$ across the fitting range is at the percent level. We quote the uncertainty stemming from this fit in the right panel of Fig.~\ref{fig:luminosity}, though the error bars are too small to be visible in the figure. This panel shows $\mathcal{L}_{\rm peak}^{\infty}$ for simulations with different $\epsilon$. When $\epsilon \ll 1$, the peak luminosity scales as
$\mathcal{L} \propto \epsilon^2$. Near criticality, it grows faster than this, reaching $\mathcal{L}_{\rm peak}^{\infty} \simeq 0.068\, \mathcal{L}_{P}
\simeq 2.5 \times 10^{51}\,\mathrm{W}$ for the near-critical run of Fig.~\ref{fig:near_critical}. This is within a factor of $\sim 3$ of the
$\approx 0.2\, \mathcal{L}_{P}$ measured for critical
collapse in~\cite{Cardoso:2018nkg}, and comfortably below
the conjectured bound $\mathcal{L} \lesssim \mathcal{L}_{P}$. Beyond criticality, the peak luminosity decreases, as part of the outgoing flux is swallowed by the newly formed BH.

\end{document}